# Defining Temperatures of Granular Powders Analogously with Thermodynamics to Understand the Jamming Phenomena

Tian Hao

*Nutrilite Health Institute*
*5600 Beach Boulevard, Buena Park, CA 90621, USA*

**Abstract**

For the purpose of applying laws or principles originated from thermal systems to granular athermal systems, we may need to properly define the critical "temperature" concept in granular powders. The conventional environmental temperature in thermal systems is too weak to drive movements of particles in granular powders and cannot function as a thermal energy indicator. For maintaining the same functionality as in thermal systems, the temperature in granular powders is defined analogously and uniformly in this article. The newly defined granular temperature is utilized to describe and explain one of the most important phenomena observed in granular powders, the jamming transition, by introducing jamming temperature and jamming volume fraction concepts. The predictions from the equations of the jamming volume fractions for several cases like granular powders under shear or vibration are in line with experimental observations and empirical solutions in powder handlings. The goal of this article is to lay a foundation for establishing similar concepts in granular powders, allowing granular powders to be described with common laws or principles we are familiar with in thermal systems. Our intention is to build a bridge between thermal systems and granular powders to account for many similarities already found between these two systems.

## 1. Introduction

As we already know, thermal energy can drive an atom or a molecule's movement in gases, liquids and solids. In colloidal suspensions where small particles are dispersed in a liquid medium, thermal energy can also drive particles movement too; such a phenomenon is called the Brownian motion if the particle size is smaller than 1 micron. For granular powder systems where there is no a dispersing medium except air staying in the interstitial spaces between particles, thermal energy usually is too weak to move the particles, making any contributions to particle movements negligible. This is one of the big differences between conventional thermal systems and granular powder systems. However, there are many articles demonstrating both experimentally and theoretically that granular materials behave like molecular thermal systems [1] [2] [3] [4] [5] [6] [7]. In the article titled "Theory of Powders", Edwards [2] formulated a theory of granular powders anologically with the statistical mechanics and transport theories of regular thermal systems, and introduced the "compactivity" concept of a similar functionality of the temperature in thermodynamics. This approach was futher extended to powder mixtures where the statistical mechanics was applied to map out phase separations [1], phase diagrams [8,9], jamming transition and mixing separation [8,9]. In Edwards' analogous statistical mechanical approach, the roles of energy traditionally played in thermal systems were replaced by the free volume per particle in granular powders, which was found to be capable of predicting phase diagrams of jammed granular matter [9] and agreed with the experimental results [10]. For example, the force fluctuations in packed beads were experimentally found to obey a simple exponential law [11] and can be elegantly predicted with similar Edwards' approaches [12,13]. The extended stress ensemble mirroring the equilibrium statistical mechanics was well applied to the deformable grains [14,15] for addressing particle packings and jamming transitions, with experimental confirmation [16]. Not only the stress but also the force-tile area were argued to play an important role in addressing the stress distribution [17,18], though an angularly anisotropic orientation correlation was experimentally found to be critical, too [19]. Clearly, both the experimental and theoretical evidences suggest that granular powders can be analogously treated with the principles or laws extracted from thermal systems, though the traditional temperature concept should be modified accordingly for granular powders.

In thermodynamics, the temperature may be expressed as:

$$T = \frac{\partial E}{\partial S} \quad (1)$$

where $E$ is the internal energy, and $S$ is the entropy. In the Edwards' theory, the energy was replaced by the volume actually taken by the powder, $V$, thus Edwards' granular temperature was defined as:

$$T = X = \frac{\partial V}{\partial S} \quad (2)$$

Since $X = 0$ indicates that the volume of power is not going to change with the entropy, the most compacted case, while $X = \infty$ represents the least, Edwards called this parameter as the compactivity of powders. Nonetheless, Edwards' granular temperature is not easy to be estimated due to the difficulty of obtaining the entropy dependence information; In addition, the temperature defined with Eq. (2) will acquire a different unit than the traditional temperature, not very intuitive to analogously utilize the thermodynamic laws. By constructing analogous entropy and internal virial functions in granular powders that are equivalent to the entropy and the energy in thermal systems, the granular temperature was defined very similarly to that in thermal systems as [14]:

$$T = \frac{1}{\alpha} = \frac{\partial \Gamma}{\partial S} \qquad (3)$$

again $S$ is the entropy, $\Gamma$ represents the internal virial equivalent to energy, and $\alpha$ denotes the pre-factor in front of $\Gamma$ in Boltzmann distribution. This kind of temperature definition is frequently used in nonequilibrium thermodynamic processes [20,21]. Although the temperature definitions shown in both Eq. (2) and (3) is in line with the traditional thermodynamic temperature definition shown in Eq. (1), maintaining the original meaning of temperature is still difficult, as the kinetic energy term is missing in granular powders, in contrast with that in thermal systems where the kinetic energy is always clearly associated with temperature. Experimental and numerical results have verified that this kind of temperature definition for the granular systems of slowly moving particles works [9,22,23,24].

For the granular systems of fast moving particles, the granular temperatures are usually defined in consistent with that in ideal gases using the kinetic energy connection, $\frac{3}{2} k_B T = \frac{1}{2} m v^2$, where $k_B$ is the Boltzmann constant, $m$ is the mass of the particles, and $v$ is the velocity of particles [25,26,27,28]. This kind of granular temperature definition can be easily traced back to the conventional temperature concept in thermal systems, thus the Boltzmann equation can be applied to such fast moving particle systems. Nonetheless, the distribution function is found to be reproducible, but often not Gaussian [29]. Undoubtedly, no matter which approaches are taken, the definition of granular temperature is always focused. Properly addressing the granular temperature would definitely create a bridge easily connecting the traditional thermodynamic principles to nonequilibrium even athermal systems like granular powders.

In this article, the granular temperature is defined with the second approach mentioned above, i.e., borrowing the exact ideas from traditional thermodynamics and utilizing the kinetic energy connection shown earlier. There are two reasons that this approach is preferred: First, defining granular temperature only in this manner may allow us to apply the fundamental Boltzmann distribution equation to granular powders; Second, this approach may create a simplistic route without introducing mystery parameters like entropy and internal energy, most time hard to be determined in granular systems. For avoiding any confusions and easily distinguishing granular powders from traditional thermodynamics systems, we may term the "temperature" in granular powders as the granular temperature in this article, which will be expressed as $T_{gp}$ rather than $T_g$, as the latter is frequently referred to the glass transition temperature in polymeric and ceramic materials fields. My previous articles [30,31] have demonstrated that the popular powder flowability criteria scaled with Carr index or Hausner ratio and the rich powder flow behaviors including jamming phenomena can be well understood with the aid of the properly defined granular temperatures via kinetic energy approaches for simple sheared granular powder systems. The same temperature definition approach will be further expanded to other popular granular powder systems like powders under a vibration shaker or free flowing on a slope. The ultimate goal is to find the jamming temperatures at which the granular systems start to jam in a uniform manner and thus the jamming phenomena can be well understood; the physical treatments of jamming phenomena in granular powders are thus unified with the uniformed granular temperature definitions across all popular granular powder flowing cases.

The article is arranged as follows: We first examine if the four thermodynamic laws can still hold for granular powders; We then consider several common cases of granular matter and define granular temperatures using the kinetic energy approach across all cases with an uniform and consistent manner; The important jamming phenomena are discussed right after the granular temperatures are defined; The temperatures at jamming points are extensively addressed and defined consistently by assuming that the jamming is caused by particles incapable of moving within the half distance of the inter-particle spacing available in granular powder systems. The particle volume fraction thus starts to play a role in jamming phenomena and the jamming volume fraction equations are therefore obtained by assuming that the ratio of the granular temperature to the jamming temperature equals to one. The reason behind this assumption is very simple: if the granular temperature is analogously assumed to be the environmental temperature and the jamming temperature is assumed as the "solidifying" temperature in thermal systems, granular systems start to jam when the environmental "granular temperature is equal to the jamming temperature. The predictions from the jamming volume fraction equations are qualitatively compared with the experimental observations or results available in literature; The future attempts based on the newly defined granular temperatures will be discussed and the final summary and conclusions will be presented at the end.

## 2. Theory

### 2.1 Four laws of thermodynamics

In thermodynamics, there are four laws generally applied to any thermal system [32][33]. The zeroth law of thermodynamics states that if a thermal system *A* is in thermal equilibrium with a thermal system *B* and the thermal system *B* is in thermal equilibrium with a thermal system *C*, then thermal system *A* will be in thermal equilibrium with the thermal system *C*. The underlying implication is that if we want to know two thermal systems are at the same temperature, it is unnecessary to bring those two systems together in contact to wait for equilibration; it can be told by a third temperature medium-a thermometer-that can measure the temperature. Back to granular systems, we should be able to tell if two granular systems are in equilibrium state via a granular temperature parameter defined in such a way that the granular temperature has the same functionality as the temperature in thermal systems. The first law of thermodynamics is about the conservation of energy: the change of internal energy of a closed system is equal to the change of the heat that the system adsorbed or given off plus the work that is done on the system or by the system. In other words, the energy cannot be created without the expense of other forms of energy or destroyed without the creation of other forms of energy. This should be true for granular systems, too, though many granular systems have a dissipative nature due to the interparticle frictional forces and inelastic collisions [7]. The second law of thermodynamics is about entropy that scales the degree of disorder or a randomness of a system. The entropy should increase over time in an isolated thermal system, approaching to a maximum value. In granular systems under a vibration or a shear field, the entropy should increase with time, too, as more particles would participate the movements due to interparticle interactions and continuous application of an external excitation. The third law of thermodynamics states that the absolute zero temperature is unattainable, as thermal motions never can stop. Unlike an ideal gas system, the particles in a granular powder cannot move freely without any external mechanical perturbation, if they are not aerated or cannot flow by themselves due to gravity. As we know in ideal gas systems, the gas molecules can fly around due to the thermal energy, as the weights of molecules are negligible. However, in granular systems the driving force expelling particles to move is the external mechanical force or the gravitational force from particles themselves. The driving force is zero if there is no such an external mechanical force or the particles sit quiescently, due to the cancellation of the gravitational force of particles resulted from the supporting particles that hold the particles unmovable. This by no means indicates that there is no pressure on the wall of the container and the granular temperature is zero.

Consider a granular powder sitting inside a cylinder shown in Figure 1. As indicated by Janssen's equation [34,35], the pressure on vertical direction, $P_v$, may be expressed as:

$$P_v = \frac{\rho g D}{4\mu K}\left[1 - exp\left(-\frac{4\mu K z}{D}\right)\right] \qquad (4)$$

where $\rho$ is the true density of the particle material, $g$ is the gravity constant, $D$ is the diameter of the cylinder, $\mu$ is the frictional coefficient between the particles and the wall of the cylinder, $z$ is the depth where the pressure is considered, and $K$ is the ratio of the horizontal pressure to the vertical pressure with the relationship:

$$K = \frac{P_h}{P_v} \qquad (5)$$

The pressure on the bottom of the cylinder should be:

$$P_{vb} = \frac{\rho g D}{4\mu K}\left[1 - exp\left(-\frac{4\mu K h}{D}\right)\right] \qquad (6)$$

where $h$ is the height of the powder bed. Note that the horizontal pressure at the top is equal to zero and at the bottom can be simply estimated with Eq. (5) and (6). Since the horizontal pressure is dependent on the powder depth, the average pressure may be approximately expressed as

$$P_{ha} = \frac{\rho g D}{8\mu}\left[1 - exp\left(-\frac{4\mu K h}{D}\right)\right] \qquad (7)$$

by simply adding the horizontal pressures at the top and at the bottom and then divided by two. The average pressure on the cylinder surfaces may be written:

$$P_{av} = \frac{P_{ha}+P_{vb}}{2} = \frac{\rho g D(2+K)}{16\mu K}\left[1 - exp\left(-\frac{4\mu Kh}{D}\right)\right] \qquad (8)$$

Note that this is just a simplified approach for giving readers an idea what is the newly defined granular temperature looks like, without knowing the depth or position of the particles in the bins. As the pressure is dependent on the depth, the granular temperature is not uniform and should have a gradient. I simply give an approximated average temperature for concept demonstration purpose. There are extensive publications on utilizing kinetic gas theory to treat granular powders and the theoretical treatments are aligned with experimental results [4,7,36,37,38,39,40,41,42,43,44,45,46], implying that we may be able to define granular temperature analogously with kinetic gas theory. According to the kinetic theory of gases [47], the pressure of a gas may be expressed as:

$$P = \frac{nmv_{rms}^2}{3} \qquad (9)$$

where $n$ is the number density of molecules, $n = N/V$, with $N$ is the number of the molecules, $V$ is the volume, $m$ is the mass of a molecule, $v_{rms}$ is the root-mean-square velocity. In addition, the kinetic energy of a molecule may be expressed as:

$$\frac{1}{2}mv_{rms}^2 = \frac{3}{2}k_B T \qquad (10)$$

where $k_B$ is the Boltzmann constant. Combing Eq. (9) and (10), one may obtain the relationship between the pressure and the temperature as:

$$P = nk_B T \qquad (11)$$

Eq. (11) is the ideal gas law. If one considers the pressure expressed in Eq. (8) in granular systems is caused by the imaginary particle movement, then the granular temperature may be defined similarly as:

$$T_{gp} = \frac{\rho g D(2+K)}{16n\mu K k_B}\left[1 - exp\left(-\frac{4\mu Kh}{D}\right)\right] \qquad (12)$$

Since granular temperature is defined analogously with the kinetic energy connection $\frac{3}{2}k_B T = \frac{1}{2}mv_{rms}^2$, the Boltzmann constant, $k_B$, remains the same physical meanings as in thermodynamics, i.e., a parameter scales the thermal energy in thermal systems with regular temperature and the analogous "thermal" (or kinetic) energy in granular systems with the granular temperature. Also, $n$ should be a very large number, $T_{gp}$ is expected to be very small, close to zero, which seems to be reasonable, as at such conditions there is almost no particle movement in the system. Under this temperature definition, one may claim that the absolute zero granular temperature is unattainable, even when a whole granular system is in a stationary state, which is very similar to the third law of thermodynamics. In summary, the four laws of thermodynamics may be analogically applied to granular systems with apparently different but essentially the same definition of temperatures. A comparison between thermal systems and granular powders is given in Table 1. Note that the granular temperature is defined in the same manner as the regular temperature in thermodynamics, the parameter "$Q$" in granular powders maintains the same physical meaning as in thermal systems. Again, this is the "beauty" of defining granular temperatures using the same approach as defining the regular temperature. Many parameters maintain the same meanings and the familiar thermodynamic principles can thus be applied to granular powders, which is the main reason that I prefer to use the kinetic energy connection approach to define the granular temperatures.

Table 1, Four laws of thermodynamics in thermal systems and granular powders

|  | Thermal systems | Granular powders |
|---|---|---|
| The zeroth law | If $T_A = T_B, T_B = T_C$, then $T_A = T_C$ | Same $T_{gp}^A = T_{gp}^C$ |
| The first law | Conservation of energy, $\Delta E^{tot} = Q + W$, where Q is heat and W is work. | Same $\Delta E^{tot} = Q + W$ |
| The second law | Entropy tends to increase, $\Delta S \geq 0$ | Same $\Delta S \geq 0$ |
| The third law | Absolute zero temperature is unattainable, $T \neq 0$ | Same $T_{gp} \neq 0$ |

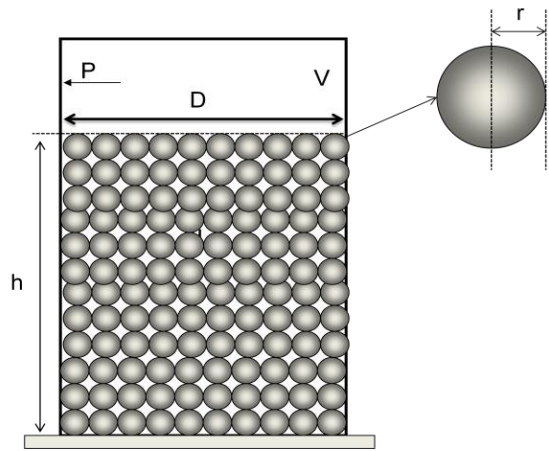

*Figure 1 A granular powder sits inside a cylinder without any movement.*

## 2.2 Granular temperatures of common powder flows and tapping processes

Now we may turn attention to define the granular temperatures of common powder flows and tapping processes. They are schematically illustrated in Figure 2: a) a powder under a simple shear; b) a powder rolling on a slope; and c) a box of a powder under a vibration. We will start with the case (*c*), as this case is relatively complicated and was already addressed in my previous publication. Let's consider a very simple granular system—a box of the volume *V* with many spheres sitting inside as shown in Figure 2 (c). Since the spheres have weights, they will generate a pressure on the bottom of the box and the sides of box, too. As shown earlier in Eq. (4), the pressures on the sides should differ from the total weight of all spheres. The whole box is fixed on a plate that can move horizontally back and forth with a vibration expressed as $L = L_0 \exp(i\omega t)$ [48], where $L_0$ is the amplitude of vibration, $\omega$ is the angular frequency, and $t$ is the time. When an external vibration is not applied to the granular system, all particles are stationary and at this moment the granular temperature is very close to zero, as indicated in Eq. (12). The entropy of the whole system should be very small, too. When an oscillatory vibration is applied as $L = L_0 \exp(i\omega t)$, the energy flow rate to the granular system may be calculated as [49]:

$$\dot{E} = F(t) \cdot v(t) \qquad (13)$$

where $F(t)$ and $v(t)$ are the force and velocity at the interface, respectively. One may assume that $F(t) = Mg$, i.e., the force is equal to the weight of whole spheres inside the box, $M = \frac{4}{3}\pi r^3 \rho n$, $g$ is again the gravity constant, $\rho$ is the true density of particles. The term $v(t)$ may be assumed as a constant, expressed as the amplitude divided by the time within a cycle:

$$v(t) = \frac{L_0}{1/(\omega/2\pi)} = \frac{L_0 \omega}{2\pi} \qquad (14)$$

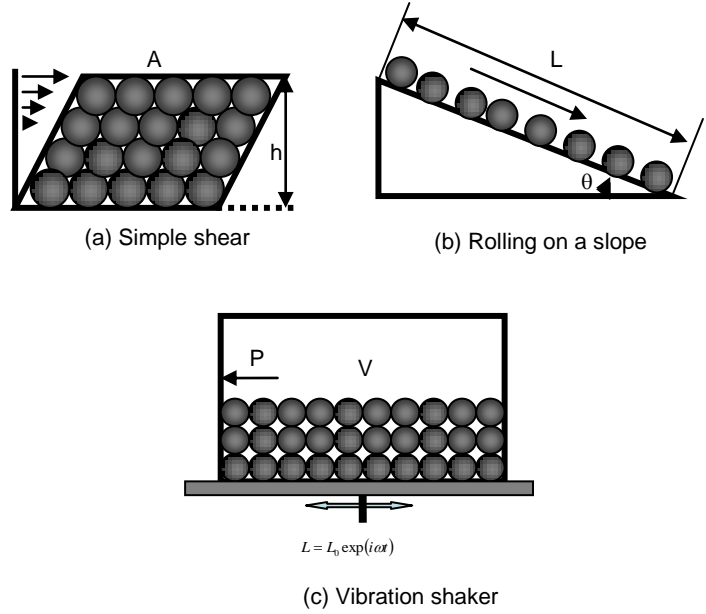

*Figure 2 Granular systems under (a) a simple shear; (b) rolling by themselves on a slope; (c) under a horizontal vibration $L = L_0 \exp(i\omega t)$.*

Please note that the vibration shaker discussed in the article is different from the regular mechanical vibration with a spring. If it is connected with a spring, the velocity should be time dependent, as the movements have to follow the Hook's law. However, in the vibration shaker case, since there is no a spring attached, the velocity is thus assumed to be a constant, which is very close to the real situation in vibration shaker experiments. The energy flow rate from the vibration shaker to the granular system is thus expressed as:

$$\dot{E} = \frac{MgL_0\omega}{2\pi} \qquad (15)$$

If the number of vibrations is assumed to be $n_v$, then the total time spent in vibration may be expressed below:

$$t = n_v \times \left(\frac{1}{\omega/2\pi}\right) = \frac{2\pi n_v}{\omega} \qquad (16)$$

The total energy flowing into the powder system may be expressed as:

$$E = \dot{E}t = MgL_0 n_v \qquad (17)$$

Eq. (17) may indicate that the total energy transferred into the powder system seems to be independent of the frequency of vibration, and only dependent on the amplitude of vibration. However, please note that Eq. (16) tells $\frac{n_v}{t} = \frac{\omega}{2\pi}$, implying that $n_v$ is dependent on the vibration frequency, thus the energy input is also dependent on the vibration frequency. According to the kinetic theory of gases [47], the kinetic energy of a molecule may be expressed as Eq. (10). If the number of molecules is $N$, then the total kinetic energy is

$$E = \frac{1}{2}mv_{rms}^2 N = \frac{3}{2}Nk_BT \qquad (18)$$

As indicated earlier, there are a large number of theoretical treatments of granular flows using the analogy of molecular fluids via standard statistical mechanics and kinetic theory, which are generally in a good agreement with the experimental results [4 7 36 37 38 39 40 41 42 43 44 45 46 50 51 52 53 54]. We thus continue to utilize the kinetic theory to analogously define the granular temperatures. Assume that the energy flowing to the granular system contributes to the movement of particles inside the box. Replacing the temperature in Eq. (18) with the granular temperature, one may easily reach

$$E = MgL_0 n_v = \frac{3}{2}Nk_B T_{gp} \qquad (19)$$

If the particles have the true density of $\rho$ and radius $r$, then

$$M = N\frac{4}{3}\pi r^3 \rho \qquad (20)$$

Substituting Eq. (20) into Eq. (19) and re-arranging may lead to the granular temperature of particles under a vibration:

$$T_{gp} = \frac{8}{9} \frac{\pi r^3 \rho g L_0 n_v}{k_B} \qquad (21)$$

The granular temperature defined in Eq. (21) has a unit of Kelvin, same as the regular temperature for thermal systems. For one micrometer sized particles of density 1 $g/cm^3$ under a vibration, $L_o = 1\ cm$, $n_v = 1000$, the granular temperature expressed in Eq. (21) is equal to $1.98 \times 10^{10}\ K$, a very high temperature in comparison with the temperature in thermal systems, However, in thermal systems the molecules or particles usually travel in sub-micrometer scaled distances, while in vibrated granular powder systems particles may travel in a full distance of the vibration amplitude, a centimeter scaled distance. The traveling distance difference between those two movements is approximately in the order of $10^5 \sim 10^7$, which makes the granular temperature relatively on par with the conventional thermal temperature.

Note that the temperature defined above is only appropriate for granular systems with an external vibration excitation. If a granular system is under a simple shear as shown in Figure 2(a), the granular temperature should be defined differently, as the energy flowing into the granular system is different. Suppose that the shear stress is σ and the shear rate is $\dot{\gamma}$, then the force F and the velocity v may be expressed as:

$$F = \sigma A, \qquad v = \dot{\gamma} h \qquad (22)$$

where $A$ is the area of the sample and $h$ is the thickness of the sample. On the basis of Eq. (13), the injected energy flowing rate from a simple shear field may be expressed as:

$$\dot{E} = Fv = \sigma \dot{\gamma} A h = \sigma \dot{\gamma} V \qquad (23)$$

where $V$ is the volume of the granular system, $V = M/\rho_b = \frac{4}{3}\pi r^3 N \rho / \rho_b$, $\rho$ and $\rho_b$ are the true and bulk density of the granular powder, respectively. Using Eq. (19), the granular temperature of a sheared powder after a time period of $t$ may be expressed as:

$$T_{gp} = \frac{8}{9} \frac{\pi \sigma r^3 \dot{\gamma} \rho t}{k_B \rho_b} \qquad (24)$$

Please note if the "granular temperature" is defined as the mean squared grain speed fluctuation, rather than the kinetic energy connection used in this article, the obtained granular temperature would be scaled with the square of shear rate [55], rather than the linear relationship shown in Eq. (24). Since the mean squared grain speed fluctuation is hard to estimate in granular systems and the total energy input is easy to obtain, I prefer to define the granular temperatures via energy input. Please keep in mind that the kinetic energy $E = \frac{1}{2}mv^2$, clearly shows there is a "square" difference between energy and speed, ultimately resulting in different shear rate dependencies of the defined granular temperatures; In my definition, the granular temperature should be linearly dependent on shear rate. Again, assuming one micrometer sized particles of true density 1 $g/cm^3$, bulk density 0.3 $g/cm^3$, under a shear field, $\sigma = 1\ Pa$, $\dot{\gamma} = 1\ Hz$, and shearing for 5 min., $t = 300\ s$, the granular temperature expressed in Eq. (24) is equal to $2.02 \times 10^8$ K, still a very high temperature.

If granular spheres flow over a slope as shown in Figure 2 (b), the granular temperature should be defined differently, too. The force that drives spheres to move downward should be $mg\sin\theta$, where $m$ is the mass of a sphere, $\theta$ is the angle of slope. If the friction coefficient between the particles and the slope surface is $\mu$, the frictional force should be $\mu mg\cos\theta$. The net force on a particle may be expressed as:

$$F = mg\sin\theta - \mu mg\cos\theta \qquad (25)$$

According to Newton's second law, $F = ma$, where $a$ is acceleration, one may find

$$a = g(\sin\theta - \mu\cos\theta) \qquad (26)$$

The initial velocity of a sphere at the top of the slope is zero and at the time $t$ the velocity is assumed to be $v$, thus

$$\frac{dv}{dt} = a \qquad (27)$$

which is the definition of acceleration. Using the energy defined in Eq. (17), i.e., the energy is the energy rate multiplied by the time, leads to:

$$E_p = Fvt = mg^2(\sin\theta - \mu\cos\theta)^2 t^2 \qquad (28)$$

Eq. (28) gives the energy of one single particle. For a granular powder containing $N$ particles, the total energy may be expressed as:

$$E = E_p N = Nmg^2(\sin\theta - \mu\cos\theta)^2 t^2 \qquad (29)$$

Using Eq. (19) again, one may obtain granular temperature for spheres on a slope

$$T_{gp} = \frac{2}{3k_B} mg^2(\sin\theta - \mu\cos\theta)^2 t^2 \qquad (30)$$

For particles of radius $r$, $m = \frac{4}{3}\pi r^3 \rho$, so Eq. (30) may be further written as:

$$T_{gp} = \frac{8\pi r^3 \rho}{9k_B} g^2 (\sin\theta - \mu\cos\theta)^2 t^2 \qquad (31)$$

Assuming one micrometer sized particles of true density 1 $g/cm^3$, $\theta = 45°$, $t = 10\ s$, $\mu = 0.3$, then the granular temperature defined in Eq. (31) is equal to $4.76\times10^{11}$ K, even much higher temperature than the one defined in vibration conditions. Note that this definition is suitable for idealized conditions where there is no interparticle collisions and the particle can move freely under the gravitational force, which leads to a very high granular temperature. The actual net force and acceleration should be much smaller than that indicated in Eq. (25) and (26). If there is a 10% reduction in both force and acceleration due to the resistance from other particles, the obtained granular temperature would be 100 times smaller, which sets the granular temperature on par with that defined in vibration and shear cases.

### 3. Particle jamming and associated temperatures

Jamming is a very common phenomenon in granular powders, where particles suddenly stop moving due to the strong connectivity or interaction between particles in a constrained space [9,56,57,58,59]. There are two kinds of jamming phenomena: static jamming occurring in dense systems due to the spatial congestions and dynamic jamming due to the complications and competencies between the shearing and crowdedness in confined spaces. The jamming phenomena are very similar to the first order phase transition observed in thermal systems, where a liquid state transitions to a solid state due to the temperature drop, and the whole system changes from a free flow state to a solidified stationary state [59,28,60]. However, physically the jamming acts more like the second order phase transitions such as glass transitions and percolation transitions. In this article, the jamming is defined as the immobility of particles due to the low kinetic energy (low granular temperature) and the spatial crowdedness (the very small free volume) in the systems. It would be interesting to evaluate the granular temperatures at jamming points based on the definitions proposed earlier. Since the granular temperature attains a very similar functionality as the conventional temperature, we may analogously assume that the "thermal" energy from the granular temperature is the source of particle motions and thus associated with particle jamming, too. The jamming will be defined as a phenomenon when particles are unable to travel the allowed free distance on the basis of free volume in a granular system. The interparticle spacing (IPS) of a granular powder may be expressed as [61,62]:

$$\text{IPS} = 2\left(\sqrt[3]{\emptyset_m/\emptyset} - 1\right)r \qquad (32)$$

where $\emptyset_m$ is the maximum packing fraction, $\emptyset$ is the particle volume fraction, and r is the particle radius. At a free flowing un-jammed state, particles are supposed to have the energy capable of travelling the full distance shown in Eq. (32). However, at jammed states, particles don't have sufficient energy and are assumed to be capable of "vibrating" within the half of the distance expressed above. Note that the IPS equation above is derived on the basis of Kuwabara's cell model [63] and the half the IPS distance means that there is a great extent of virtual cell overlap between two particles, implying that these two particles touch each other. Under such a definition of jamming state, the energy required for *N* particles to move a half the IPS distance may be expressed as:

$$E = Nmg\left(\sqrt[3]{\emptyset_m/\emptyset_j} - 1\right)r \qquad (33)$$

where $\emptyset_j$ is the particle volume fraction when particles are jammed. According to Eq. (18), the energy shown in Eq. (33) should be equal to the kinetic energy for particles, which has been used in this article many times for defining the granular temperatures. Therefore, the granular temperature at jamming points may be expressed as:

$$T_J = \frac{2mg\left(\sqrt[3]{\emptyset_m/\emptyset_j} - 1\right)r}{3k_B} \qquad (34)$$

where $T_J$ is the granular temperature at a jamming point. Since $m = \frac{4}{3}\pi r^3 \rho$ by definition, Eq. (34) may be further written as:

$$T_J = \frac{8\pi\rho g\left(\sqrt[3]{\emptyset_m/\emptyset_j} - 1\right)r^4}{9k_B} \qquad (35)$$

Since $\emptyset_m$ is only related to the packing structure for mono-dispersed particle systems [56,64], one may infer that jamming transition temperature is dependent on the true density of the particulate materials, the radius of the particles, and the particle volume fractions at jamming points. For obtaining an intuitive idea how jamming transition temperatures change with the particle volume fractions at jamming points, we schematically plot Eq. (35) against both particle volume fraction at jamming points and particle radii in Figure 3, under assumption that $\rho=1$ *g/cm³* for typical true density of polymer materials, $\emptyset_m = 0.74$ for cubic or hexagonal close-packed systems. The jamming temperature generally decreases with the increase of particle volume fractions at jamming points and are strongly dependent on the particle sizes. There are several orders of magnitude difference among the jamming temperatures when the particle sizes only increase 10 times. This is probably due to the fact that the jamming temperatures are directly proportional to the 4$^{th}$ powers of the particle radius as indicated in Eq. (35). When the particle volume fractions at jamming points approach the maximum volume fraction, the jamming temperatures quickly drop to a very low temperature, regardless if the particle sizes are large or small. Such low jamming temperatures thus imply that whole granular systems are fully jammed and solidified. We may analogously call them "frozen points" as observed in thermal systems, and every particles are locked at certain sites without any movement. This is the beauty of defining granular temperature analogously and consistently using the kinetic energy connections, where both granular temperature and traditional temperature attain a similar physical meaning and thus granular temperature is easily to be comprehended. Another apparent benefit is that we may possibly employ the theoretical framework like Eyring's rate process theory and free volume concept originated from thermal systems to treat granular powders, unifying both systems with a single approach. In addition, Fig.3 shows small particles tend to jam at very low granular temperatures, which is understandable and consistent with practical observations: large granules usually flow much better than small particles. When particles become smaller and smaller, interparticle forces become more important and particles tend to aggregate or bridge very easily, resulting in very poor flowability. As one may tell, for particles of radius about 0.1 micron, the jamming transition temperature is little below one Kelvin, an extremely low temperature.

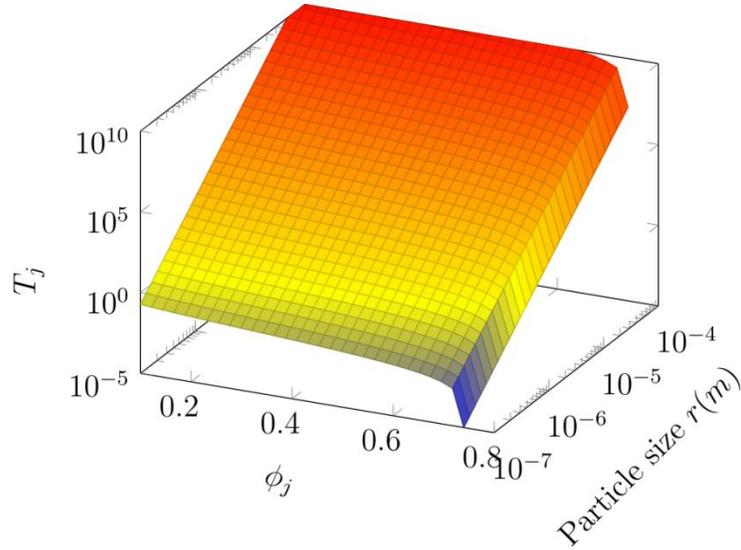

*Figure 3, Jamming transition temperature vs. the jamming volume fractions and particle radii based on Eq. (35), under assumption that ρ=1 g/cm³, $Ø_m$ = 0.74.*

While in comparison with 10 micron sized particles the jamming happens at a very high temperature, about $10^7$ K. This mainly results from the fact that the jamming temperature is defined if the external energy that drives particles to move can overcome the weight of particles over a certain distance in granular systems. This prediction seems to qualitatively agree with the experimental observation on the superheating phenomena of monodispersed metal beads of diameter 3.15 *mm* reported in literature [6]: Under a vigorous vertical shaking, a hexagonal closed packed crystal structure was observed and eventually melted away (or called evaporate in the literature) after a period of time. As stated earlier, a granular powder under a vibration may have a very high granular temperature. Based on Eq. (21), the granular temperature of such a metal beads system is in the order of $10^{20}$ K, and also, it is time dependent. Longer time vibrations will create higher granular temperatures, which could be the reason that the crystal structure was finally evaporated after a relatively long vibration. For particles of size about 0.1 micron, jamming should happen at much lower granular temperatures about one Kelvin based on Fig. 3. Such a low granular temperature may correspond to a quiescent state where no apparent motions are obviously detected. In reality, submicron or nanometer sized powders of low densities typically tend to have a very poor flowability and easily form arching structures [65,66]. The newly defined granular temperatures seem to agree well with the empirical observations.

It would be valuable to explore at what conditions jamming could happen by simply using the granular temperatures defined earlier at several common cases divided by the jamming granular temperature defined in Eq. (35). The ratio equal to one gives the jamming conditions for particular granular systems. For a simple shear case shown in Figure 2 (a):

$$\frac{T_{gp}}{T_J} = \frac{\sigma \dot{\gamma} t}{(\sqrt[3]{\emptyset_m/\emptyset_j} - 1) g r \rho_b} = 1 \qquad (36)$$

Thus one may easily get:

$$\emptyset_j = \frac{(gr\rho_b)^3 \emptyset_m}{(gr\rho_b + \sigma\dot{\gamma}t)^3} = \frac{\emptyset_m}{[1+\sigma\dot{\gamma}t/(gr\rho_b)]^3} \qquad (37)$$

Eq. (37) defines the conditions that the jamming happens at a simple shear case. It clearly tells that the particle volume fractions at the jamming points are dependent on the shear stress, shear rate, and surprisingly the radius of the particles. For illustrative purpose, $\emptyset_j$ is plotted against $\dot{\gamma}$ and particle size over a wide range and shown in Figure 4, under the assumption that $\emptyset_m = 0.74$, $\rho_b = 0.3 g/cm^3$, $t=60$ s, $\sigma = 1$ Pa. Note that there are two regions where $\emptyset_j$ is insensitive to shear rates, very high shear rates above $10^1$ s$^{-1}$ (blue area) and very low shear rates below $10^{-5}$ s$^{-1}$ (red area). In the shear rate regime between $10^{-5}$-$10^1$ s$^{-1}$, $\emptyset_j$ dramatically increases with the decrease of shear rates. In other word, the jamming may happen at lower particle volume fractions when the shear rate increases, which is consistent with experimental observations [57,67]. When the shear rate is smaller than $10^{-5}$ s$^{-1}$, the system may only jam at high particle volume fractions; when the shear rate is about $10^{-5}$ s$^{-1}$, the volume fractions at jammed points become lower and lower under higher and higher shear rates. The system starts to completely jam when the shear rate is about $10^1$ s$^{-1}$. The big drop between $10^{-5}$-$10^{-2}$ s$^{-1}$ may indicate that a shear induced structure change happens in this area. These two regions are very similar to the "fragile states" and "shear-jammed states" observed experimentally [57,67], where the "fragile states" correspond to a strong network structure percolated in one direction and the "shear-jammed states" correspond to a strong network percolated in all directions. Back to Figure 4, the "fragile states" is somewhat similar to the big fall region between $10^{-5}$-$10^{1}$ s$^{-1}$, while "shear-jammed states" is the region where the shear rate is above $10^{1}$ s$^{-1}$. The qualitative agreement with the experimental observation may imply that the granular temperature defined in a consistent manner with the conventional temperature in thermal systems actually works.

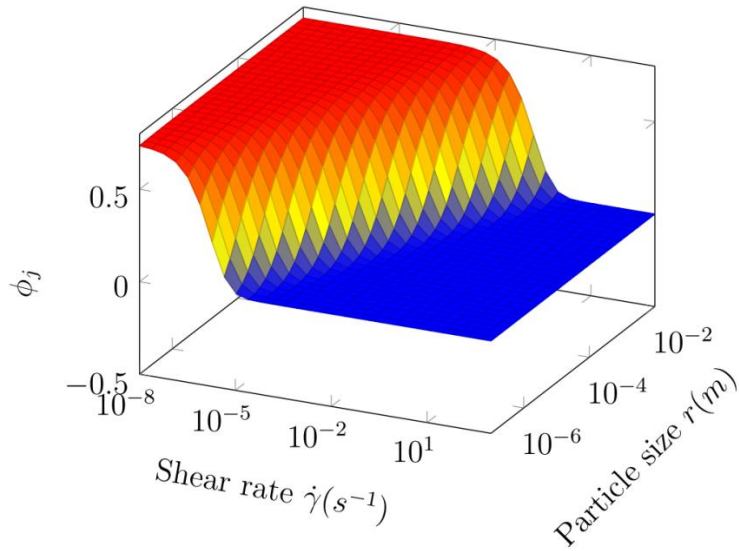

*Figure 4, The particle volume fraction at the jamming points, $\emptyset_j$, is plotted against the shear rate, $\dot{\gamma}$, and particle size, r, from Eq. (37) under the assumption that $\emptyset_m = 0.74$, $\rho_b = 0.3 \, g/cm^3$, $t=60$ s, $\sigma = 1$ Pa.*

The particle radius has a clear impact on the jammed volume fraction, too, based on Eq. (37) and Figure 4. Smaller particles can only jam at lower shear rates and melt at higher shear rates; Large particles won't jam at lower shear rates, unless the particle volume fractions are close to the maximum packing fraction. According to Eq. (37), the jamming volume fraction can be very small when the shear rate is very high, but never can be zero. In stationary conditions, the jamming can only occur in dense systems. However, Eq. (37) is again for dynamic systems under a continuous shear rather than stationary systems. What Fig.4 tells us is that jamming may occur at low particle volume fractions if shear rates are high, which is quite similar to shear thickening phenomena observed in colloidal suspension systems. When the particles have a size in micron ranges, the jamming can happen in a very wide shear rate range from $10^{-6}$ to $10^{2}$ s$^{-1}$ no matter what particle sizes are. In other words, high shear rates would possibly induce jamming easily. When particle volume fractions are close to the maximum packing fraction, particles may always jam no matter that they have a large or small particle size. Since there are two regions insensitive to shear rates, shear rates may not be a dominate factor in jamming process.

Shear stress may play a critical role in jamming process. The particle volume fraction at the jamming points, $\emptyset_j$, is plotted in Figure 5 against shear stress and particle radius, $r$, under the assumption that $\emptyset_m = 0.72$, $\rho_b = 0.3$ g/cm$^3$, t= 60 s, and $\dot{\gamma}$= 0.5 s$^{-1}$. As one may tell, particle volume fractions at jamming points are strongly dependent on both shear stress and particle size. At low shear stress and large particle size regions, granular systems may only jam at very high particle volume fractions close to the maximum packing fraction; with the increase of shear stress, granular systems may jam at lower and lower volume fractions, implying that the shear induced thickening phenomena occurs, if shear stress is high enough. Comparing Figure 5 with Figure 4, one may easily find that, at very low shear rates about $10^{-8}$ s$^{-1}$, granular systems are only going to jam when particle volume fractions are close to the maximum packing faction, no matter what the particle sizes are. This seems to contradict with what is shown in Figure 3, where jamming temperatures are strongly related to particle sizes. The discrepancy may result from low shear rates about $10^{-8}$ s$^{-1}$ and especially, low shear stress 1 $Pa$ assumed in the calculation. In reality, we may be unable to apply such a weak shear field to granular systems and anticipate to drive particle movements. In contrast, particle volume fractions at jamming points are continuously changing with particle sizes in Figure 5 under a wide shear stress range. Although large particle sizes correspond to high jamming temperature, a strong shear stress could "cool down" granular systems and thus they can jam at relatively high granular temperature based on Eq. (35) and (37). For better demonstrating the roles of both shear stress and shear rate in jamming process, the particle volume fractions at jamming points are plotted against both shear stress and shear rate in Figure 6. At very high shear stress like $10^{3}$ $Pa$, shear rate seems unable to control when jamming is going to happen and everything is dominated by the shear stress: granular systems could jamming at very low particle volume fractions irrelevant with shear rate; at very low shear stress like $10^{-3}$ s$^{-1}$, there are two distinctive regions against shear rates:

jamming at either low or high shear rates but independent of shear rate, and a narrow transition "fragile" region in the middle. Clearly, one may reach a conclusion that jamming actually is mainly dominated by the shear stress, rather than shear rate, which is consistent with experimental observations[57,68].

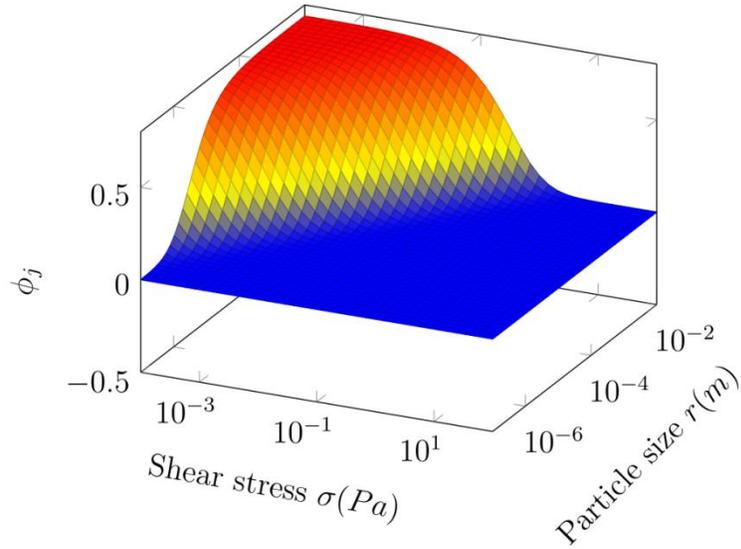

*Figure 5, The particle volume fraction at the jamming points, $\phi_j$, is plotted against shear stress σ and the particle radius, r, obtained from Eq. (37) under the assumption that $\phi_m = 0.74$, $\rho_b = 0.3 \, g/cm^3$, t=60 s, and $\dot{\gamma} = 0.5$ s$^{-1}$.*

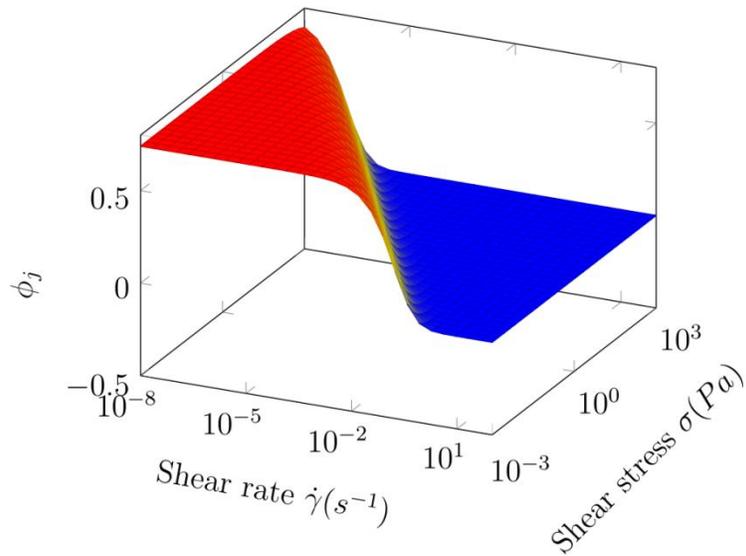

*Figure 6, The particle volume fraction at the jamming points, $\emptyset_j$, is plotted against shear stress $\sigma$ and shear rate, obtained from Eq. (37) under the assumption that $\emptyset_m = 0.74$, $\rho_b = 0.3\ g/cm^3$, t=60 s, and particle size $r = 10^{-5}$ m.*

Similarly, one may find the jamming volume fraction conditions for granular powders under a vibration. Replacing $n_v$ with Eq. (16), using Eq. (21) divided by Eq. (35) and assuming that it equals to 1 leads:

$$\frac{T_{gp}}{T_J} = \frac{L_o\ \omega t}{2\pi r(\sqrt[3]{\emptyset_m/\emptyset_j}\ -1)} = 1 \qquad (38)$$

Thus the particle volume fraction at jamming points may be expressed as:

$$\emptyset_j = \frac{\emptyset_m}{[1+L_o\ \omega t/(2\pi r)]^3} \qquad (39)$$

Eq. (39) indicates that the particle jamming volume fractions are dependent on the amplitude and frequency of the vibration, the time, and the particle radius. For clearly illustrating the relationship among $\emptyset_j$, vibration amplitude, particle radius, these parameters are plotted in

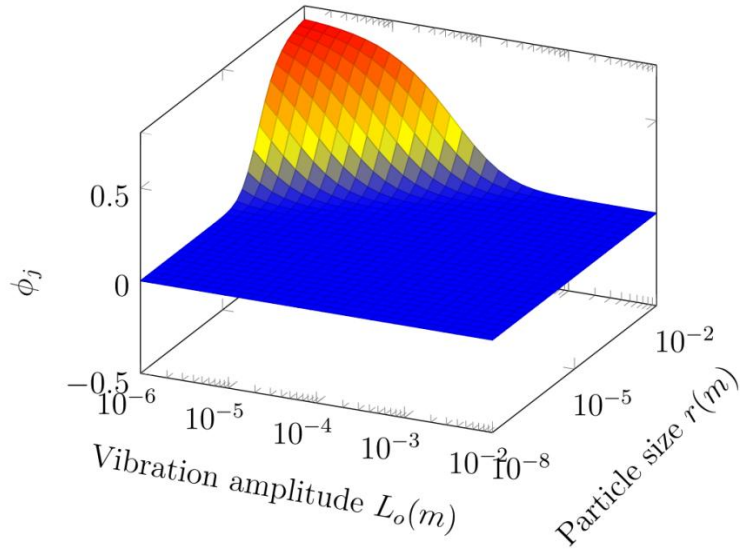

*Figure 7, The particle volume fraction at the jamming points, $\emptyset_j$, is plotted against the particle radius, r, and vibration amplitudes from $10^{-2}$ to $10^{-6}$ m, obtained from Eq. (39) under the assumption that $\emptyset_m = 0.74$, $\omega = 6.8\ rad/s$, t=60 s.*

Figure 7 based on Eq. (39) under the assumption that $\emptyset_m = 0.74$, $\omega = 6.8$ rad/s, t=60 s. Particles tend to easily jam at very low particle volume fraction at most regions, the blue area;

Particles of larger sizes tend to jam at very low vibration amplitudes if particle volume fractions are high enough, close to maximum packing fraction. Comparing Figure 7 with Figure 5, one may come to the conclusion that vibration amplitude performs similarly as shear stress in shearing case. Consider a granular system initially unjammed due to low particle volume fractions below jamming points: when the vibration amplitude increases, particles tend to jam at lower particle volume fractions and the system would finally reach the jamming point; further increase of vibration amplitude would keep the system jammed all the time, unless vibration can cause particles be dispersed in a larger space and the particle volume fraction is lowered. Once the particle volume fraction become below the jamming volume fraction, the jammed particles start to melt and would jam again under further increase of vibration amplitudes. This "melt-jam" meta-stable region could span a large region as demonstrated in Figure 7. Since the vibration frequency stays at the identical position as the vibration amplitude in Eq. (39), a very similar dependency of the jamming particle volume fractions on the vibration frequency is expected, i.e., high vibration frequencies may melt the jammed particles, too. In industries, vibration conveyors with controllable amplitudes and frequencies are frequently employed to transport granular powder materials. The results expressed in Eq. (39) and demonstrated in Figures 7 seem to be consistent with the empirical practical solutions of transporting powders that have been used for a long time. For clearly demonstrating the role of vibration frequency, particle volume fractions at jamming points is plotted in Figure 8 against vibration amplitude and frequency under assumption that the assumption that $\emptyset_m = 0.74$, r $= 10^{-3}$ m, t=60 s. Vibration amplitudes cannot effectively lower jamming volume fraction until they are high enough, more than $10^{-3}$ m in Figure 8. In contrast, vibration frequency seems to have more pronounced impact on the jamming volume fractions, bringing it down to a lower volume fraction even at low frequency; the effect is even more dramatic when vibration amplitudes are high. However, note that these two distinctive regimes are independent of shear rate: in both red and blue areas, the particle volume fractions at jamming points don't change with shear rate, and the change only happens in transition regime of shear rate dependence. Vibration frequency functions comparably to shear rate, as demonstrated clearly in Figures 4 and 8. Again, jamming seems strongly dependent on vibration amplitude and the particle radii, which is similar to what is demonstrated for granular powders under shear.

It should be interesting to see how the vibration time would impact the jamming particle volume fractions, as longer time means more energy flowing into the systems, as indicated in Eq. (21). Figure 9 shows the particle volume fractions at the jamming points, $\emptyset_j$, plotted against the vibration time and the vibration amplitudes from $10^{-6}$ to $10^{-2}$ m, obtained from Eq. (39) under the assumption that $\emptyset_m = 0.74$, $\omega = 6.8$ rad/s, r$= 10^{-3}$ m. Again, the particles may jam at the very beginning of the vibration when particle volume fractions are high enough. If particle volume fraction is much lower than the maximum packing fraction at the beginning, the system may remain at unjammed state, but quickly enter into "melt-jam" regime with longer vibration time. At lower vibration amplitudes, the system remains unjammed for a relative long time, while at high vibration amplitudes, jamming may happen in much faster paths at lower particle volume fractions. Again, particle sizes could play a role in determining where the jamming points are. Figure 10 shows particle volume fractions at the jamming points, $\emptyset_j$, plotted against the vibration time and particle radii from $10^{-6}$ to $10^{-2}$ m, obtained from Eq. (39) under the assumption that $\emptyset_m = 0.74$, $\omega = 6.8$ rad/s, $L_0 = 10^{-4}$ m. At such a small vibration of amplitude $10^{-4}$ m, particle sizes are critical: at the beginning of vibration, jamming occurs at low particle volume fractions when particles are small, or at high particle volume fraction when particles are large. Even for large particles, jamming still may happen at low particle volume fractions after vibrated for a long time. If there is no change on particle volume fraction during vibrations, granular systems eventually would remain at jammed state; however, in a long time vibration particles either pack more tightly of a high particle volume fraction at gentle horizontal vibration condition, or tend to take more space of a low particle volume fraction at wild vertical vibration condition. The former will lead to jamming and the latter will lead to evaporation at the end. The "melt-jam" process would dominate in between, and jammed/melted structures may co-exist in the system. Again, those speculations are consistent with experimental observations reported in literature [6], where the crystalline structure was observed at the very beginning of a vertical vibration and quickly melt later with a continuous vibration. This issue will be addressed in detail in next section.

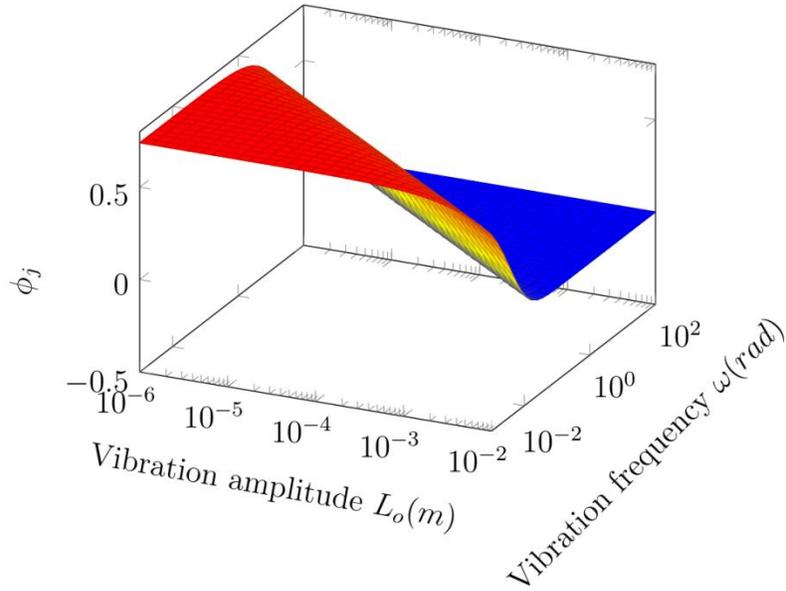

*Figure 8, The particle volume fraction at the jamming points, $\phi_j$, is plotted against the vibration amplitude and frequency, obtained from Eq. (39) under the assumption that $\phi_m = 0.74$, $r = 10^{-3}$ m, t=60 s.*

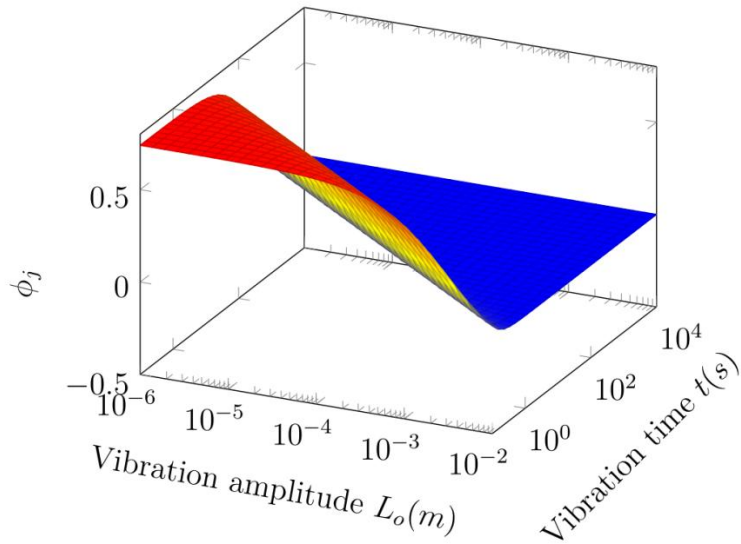

*Figure 9, The particle volume fraction at the jamming points, $\phi_j$, is plotted against the vibration time and vibration amplitudes from $10^{-6}$ to $10^{-2}$ m, obtained from Eq. (39) under the assumption that $\phi_m = 0.74$, $\omega = 6.8\ rad/s$, $r=10^{-3}$ m.*

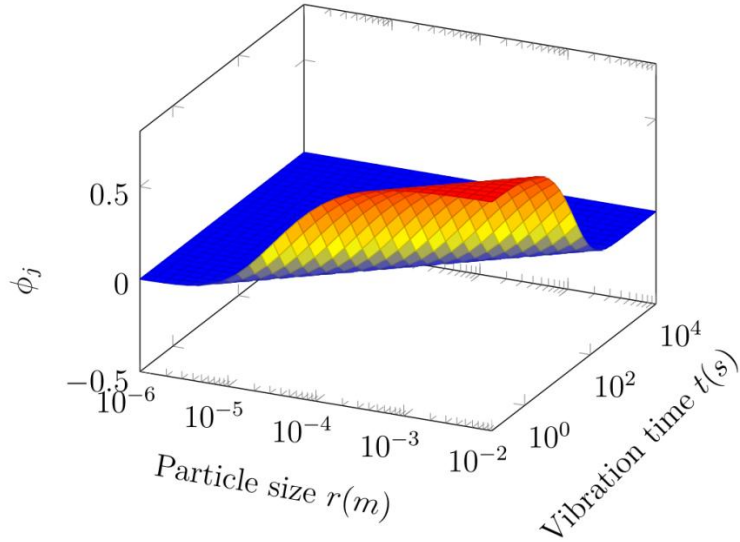

*Figure 10, The particle volume fraction at the jamming points, $\emptyset_j$, is plotted against the vibration time and particle radii from $10^{-6}$ to $10^{-2}$ m, obtained from Eq. (39) under the assumption that $\emptyset_m = 0.74$, $\omega = 6.8\ rad/s$, $L_0 = 10^{-4}$ m.*

Following the same logics and procedures described previously, one may easily obtain the particle volume fraction at jamming points for particles rolling on a slope:

$$\emptyset_j = \frac{\emptyset_m}{\left[1+\frac{gt^2}{r}(sin\theta-\mu cos\theta)\right]^3} \qquad (40)$$

and the particle volume fraction at jamming points for particles sitting inside a cylinder without any movement:

$$\emptyset_j = \frac{\emptyset_m}{\left[1+\frac{9D(2+K)}{96\mu Kr\emptyset_A}\left[1-exp\left(-\frac{4\mu Kh}{D}\right)\right]\right]^3} \qquad (41)$$

where $\emptyset_A = \frac{4\pi r^3 N}{3V}$, is the real particle volume fraction excluding all interstitial empty spaces in the cylinder, $\emptyset_A \geq \emptyset_m$. Readers are encouraged to explore the relationships among the particle volume fractions at jamming points and other related parameters under those two cases.

## 4. Experimental comparisons

Experimental evidences related to jamming process have been mentioned occasionally in previous section for a quick qualitative comparison. In this section, the predictions inferred from newly defined granular temperatures and the particle volume fractions at jamming points will be compared more intensively with experimental results available in the literature. The fundamental questions important to the jamming process will be addressed for validating the new approach employed to treat the jamming process in this article. Three popular phenomena, crystal structure evaporation under a vertical vibration, shear weakening during constant volume shearing, and shear jamming, are chosen from literature and discussed below.

First, let's turn our attention to crystal structure evaporation under vibration observed in literature [6]. In this article[6], 720 steel beads of radius $1.58 \times 10^{-3}$ m was poured on a Plexiglas hexagonal container for obtaining a hexagonal closed packed monolayer; this container was placed on a shaker to vertically vibrate for observing the packing structure change; the volume fraction of steel beads is 0.87, and the maximum packing fraction of 2D hexagonal monolayer is about 0.91, leaving a sufficient free room for steel beads to move. We will use these experimental parameters to map out the phase transitions based on Eq. (39). Figure 11 shows volume fractions at jamming points against vibration amplitude and vibration time computed with Eq. (39) with the parameters given in the literature [6]. At the beginning when both vibration amplitude and time is small, this system tends to form jammed structures; the particle volume fraction is 0.87, very close to the maximum packing fraction, 0.91. A crystallized hexagonal structure is thus anticipated, which is observed experimentally; when the vibration amplitude is about $10^{-3}$ m, the system is in "melt-jam" transition region; however, since the system is vibrated vertically without a cover, the beads may jump out of the 2D container and completely "evaporate". The evaporation starting time points at vibration amplitude $10^{-3}$ m, from Figure 10, are about several hundred milliseconds, which is again in line with the experimental observation. Figure 11 also predicts at very low vibration amplitude below about $2\times10^{-6}$ m, the hexagonal crystal structure will remain intact within the time of one second. A phase diagram against vibration amplitude and frequency is presented in Figure 4 of the literature [6]; for comparison,the volume fraction at jamming points is plotted against vibration amplitude and frequency in Figure 12, which is amazingly similar to experimental phase diagram. Let's focus on amplitude region between $10^{-4}$ ~$10^{-3}$ m that is shown in the phase diagram in literature [6]: at very low vibration frequency below about 40 Hz, the system is in "melt-jam" transition state, both "gas" and "crystal" phases co-exist in the system; Above 40 Hz, the system tends to jam all the time even at very low volume fraction, i.e., the system enters into "superheated" crystal state; with the increase of vibration amplitude, the "superheated" crystal may appear at lower frequency range below 40 Hz, as suggested in Figure 12, which is not presented in literature [6] and needs to be confirmed experimentally in the future. In a word, we have to say that our predictions agree very well with the experimental observations. Since the newly defined granular temperatures are proportional to the shearing time, it is easy to understand why the jammed crystal structure is "destroyed" and finally "evaporated" with vibration time, naturally due to higher and higher granular temperatures.

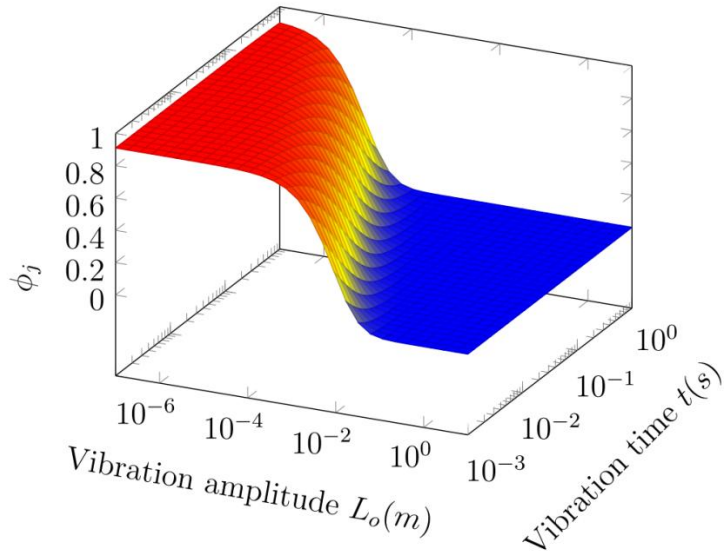

*Figure 11, The particle volume fraction at the jamming points, $\emptyset_j$, is plotted against the vibration time and vibration amplitudes from $10^{-6}$ to $10^{-2}$ m, obtained from Eq. (39) under the assumption that $\emptyset_m = 0.91$, $\omega = 376.8 \, rad/s$ or 60 Hz, $r=1.58 \times 10^{-3}$ m. All these parameters are taken from the literature [6].*

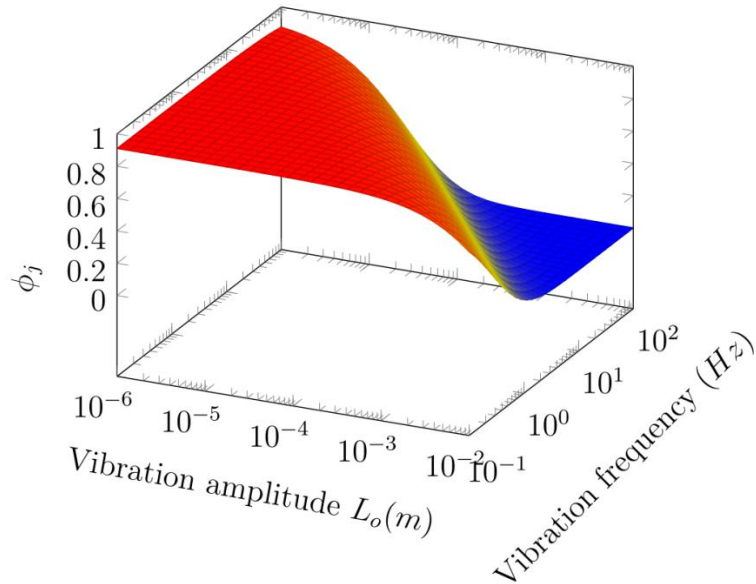

*Figure 12, The particle volume fraction at the jamming points, $\phi_j$ , is plotted against the vibration frequency from 0.1 to 100 Hz and vibration amplitudes from $10^{-6}$ to $10^{-2}$ m, obtained from Eq. (39) under the assumption that $\phi_m = 0.91$,  t $=0.2$ s,  $r=1.58\times10^{-3}$ m.  All these parameters are taken from the literature [6].*

Shear-weakening phenomena is observed in beach sand systems in a torsional shear cell [69] under constant volume shear condition. The beach sand has a bulk density 1.73 $g/cm^3$ and particle sizes ranging from 47 to 2000 micron determined with Beckman-Coulter Particle size analyzer. The sample was sheared for about 510 seconds under two conditions: constant volume and constant pressure. Shear-weakening phenomena was observed in constant volume shear condition. Although the whole system is not jammed, we may still use the volume fraction at jamming points to estimate if high shear stress is induced, under an assumption that even just partially jammed systems may generate high shear stress. Figure 13 shows the volume fraction at jamming points vs. shear rate and particle size under the assumption that $\phi_m = 0.74$, $\rho_b = 1.73$ $g/cm^3$, a typical bulk density for beach sand, t =510 s, shear stress $\sigma =10^{-2}$ Pa. These parameters are taken from the literature [69]. Take the particle sizes about $10^{-4}$ m as an example, if the sample is sheared from high to low shear rate, the sample may go through from a fully jammed state at very low particle volume fractions, a transition region of typical "jam-melt" process, and a later fully fluidic state where jamming may only happen at very high particle volume fractions that are unachievable in current beach sand systems. The transition regions could be the weakest, as the jamming may only occur at higher particle volume fractions and the system basically remains in "melting" state, as the volume is fixed and the particle volume fractions are well below jamming points. This scenario is qualitatively consistent with experimental observations: from high to low shear rate the measured shear stress steeply goes through a dip at very high shear rate regions, arises up gradually at middle shear rate regions, and levels off at low shear rate regions.

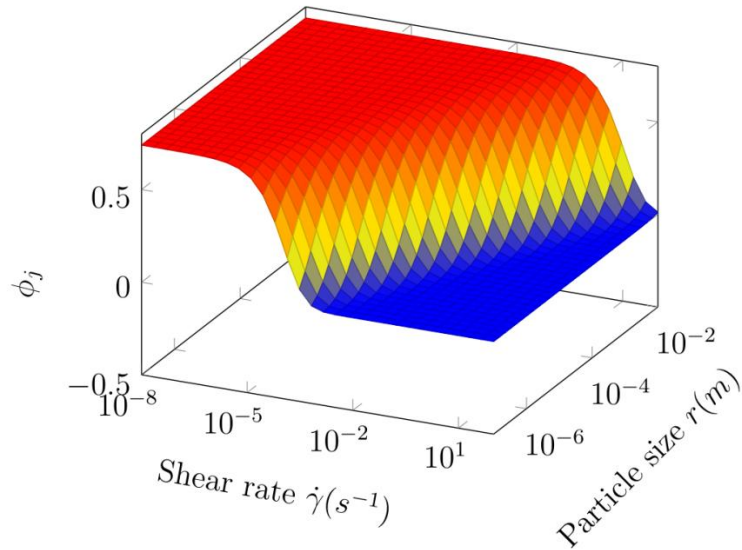

*Figure 13, The particle volume fraction at the jamming points, $\phi_j$, is plotted against the shear rate and particle size, obtained from Eq. (37) under the assumption that $\phi_m = 0.74$, $\rho_b = 1.73$ $g/cm^3$, the bulk density for beach sand, t =510 s, shear stress $\sigma =10^{-2}$ Pa. All these parameters are taken from the literature [69].*

Shear jamming phenomena were experimentally observed in suspensions of dense cornstarch particles dispersed into a density-matched solution of water, glycerol and CsCl without any change of packing fraction [68]. Let's first examine how shear rate will impact jamming phenomena. The predicted volume fraction at jamming with Eq. (37) is plotted in Figure 14 against both shear rate and particle size under a small shear stress, 10 *Pa*. The maximum packing fraction is assumed to be 0.74, the bulk density of cornstarch particles of particle size about 20 micron is 0.673 $g/cm^3$ and shearing time is only 60 *ms*, based on the experimental data shown in the literature [68]. Unless experiment can be performed at a very wide shear rate range from $10^{-8}$ to $10^2$ $s^{-1}$, jamming will be considered independent of shear rate, as only two distinctive shear regions with a very narrow transition area is predicted in Figure 14; in both distinctive regions, particle volume fractions at jamming points are independent of shear rate The "jam-melt" transition region is greatly shifted with the increase of particle sizes and therefore mostly dominated by particle sizes rather than shear rate; Shear rate could induce jamming at lower particle volume fraction, but this effect quickly diminishes with the increase of particle sizes. Therefore, it would be reasonable to say that jamming is mainly dominated by particle volume fraction and particle size. Figure 15 shows the volume fraction at jamming points vs. shear stress and particle size under conditions that $\emptyset_m = 0.74$, $\rho_b = 0.673$ g/cm$^3$, t =0.06 s, shear rate $\dot{\gamma} =10$ s$^{-1}$, which are taken on the basis of the literature [68]. As one may easily tell, predicted volume fractions at jamming points are strongly dependent on both shear stress and particle size. When shear stress goes from $10^{-2}$ to $10^5$ *Pa*, the volume fraction at jamming points slowly move from high to low, implying that shear jamming could occur at lower particle volume fractions; Unlike what we see in Figure 14 where the volume fraction at jamming points doesn't change with shear rate for several orders of magnitudes at both low and high regions, the shear stress induced jamming process is continuous until the system hits the fully jammed state. Comparing Figure 15 with the phase diagram shown in Figure 3e of the literature [68], one may find similarities between these two figures: the blue area at high shear stress and low particle size regions corresponds to fully jammed regime, the steep fragile "jam-melt" region corresponds to the "discontinuous shear thickening" regime, the less steep fragile "jam-melt" region corresponds to the "shear jamming" regime, and the narrow flat region at low shear stress and large particle size corresponds to Newtonian regime. Note that particle sizes may have a similar impact as shear stress, if particle sizes can be varied several orders of magnitudes, from $10^{-8}$ to $10^{-3}$ $m$. This is a little surprising, as intuitively jamming should be dependent on packing fraction and independent of particle sizes. Particle size dependence could be understood in this picture: in current article granular temperature is defined to be strongly related to particle size and the jamming is defined on the basis of granular temperature; The onset shear stress for inducing discontinuous shear thickening could be considered as energy need to break down the lubrication between particles, identical to the reduction of free volume available in the system, where the interparticle spacing comes to play a role and this is how the jamming temperature is defined previously. Such a prediction is consistent with experimental observation that shear jamming is induced by shear stress rather than shear rate, due to the energy or free volume requirements for granular temperature.

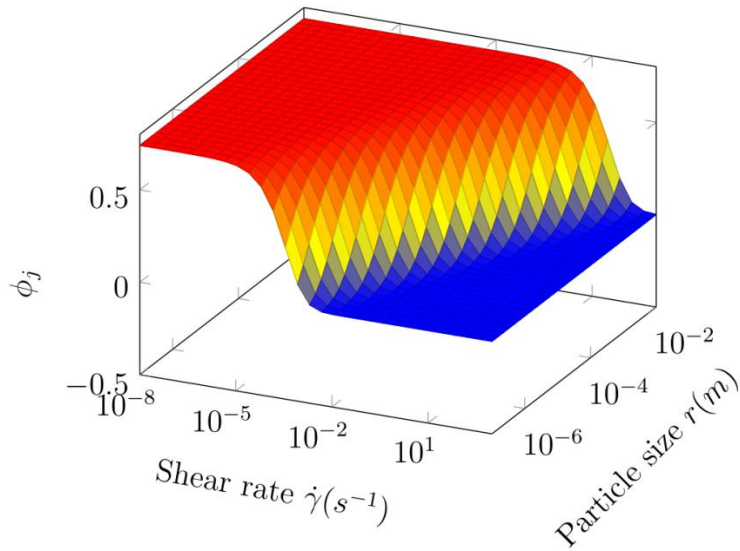

*Figure 14, The particle volume fraction at the jamming points, $\emptyset_j$, is plotted against the shear rate and particle size, obtained from Eq. (37) under the assumption that $\emptyset_m = 0.74$, $\rho_b = 0.673$ g/cm$^3$, t =0.06 s, shear stress $\sigma =10$ Pa. All these parameters are taken from the literature* [68].

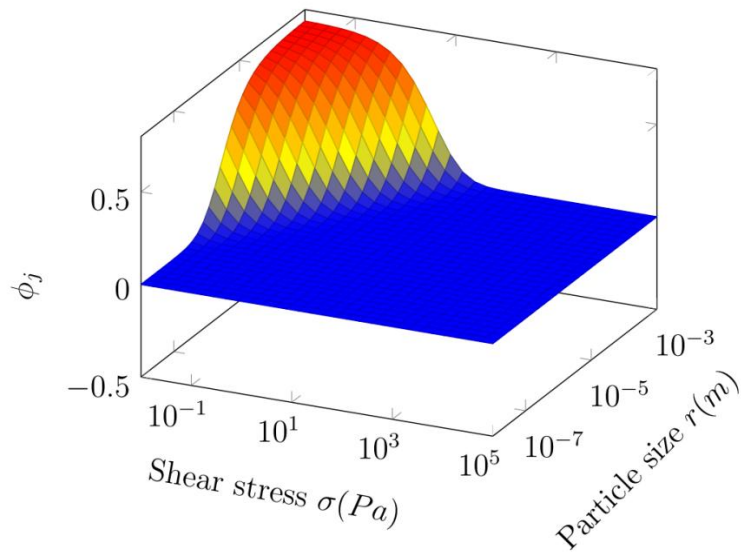

*Figure 15, The particle volume fraction at the jamming points, $\phi_j$, is plotted against the shear stress and particle size, obtained from Eq. (37) under the assumption that $\phi_m = 0.74$, $\rho_b = 0.673$ g/cm$^3$, t =0.06 s, shear rate $\dot{\gamma} = 10$ s$^{-1}$. All these parameters are taken on the basis of the literature [68].*

## 5. Discussion

I would like to emphasize that the current article simply is an extension of my two other articles published recently [30,31]. The reference [30] is a communication and the reference [31] is the full paper of that communication. As you may tell, the approach of defining granular temperatures with the kinetic energy connection $\frac{3}{2} k_B T = \frac{1}{2} m v_{rms}^2$ was already employed in these two articles for granular powders under a simple shear case, and the theoretical predictions built on this granular temperature are consistent with the empirical powder flowability criteria and even with the experimental data. Such amazing agreements prompt me to apply the same approach of defining granular temperature to other common situations like particles rolling on a slope and particles under a vibration shaker, which are the main topics of current paper. Furthermore, since the granular temperature is re-defined for granular powders, the corresponding "thermodynamics" for athermal systems is examined against the standard thermodynamics; The very common and rich jamming phenomena observed in granular powders is therefore addressed with the new granular temperature definitions for validating the new concepts and approaches proposed in this article, and most importantly providing some new insights on these complicated phenomena. Those three articles should be read through together and they gradually become much deeper and wider in sequence. Nonetheless, the approach used to define the granular temperature is same in principle across all three articles. In addition, I would like to emphasize that the approach used in these three articles to analogously define the granular temperatures is not originated by me and have been employed in many publications. I have simply borrowed the ideas from those excellent publications and extended further in my articles. Anybody questioning this approach should read the original literature first.

Among many approaches of defining granular temperatures briefed in the "Introduction" section in this manuscript, I personally think that the kinetic energy connection approach is the best, as such an approach may allow us to employ the well established thermodynamic principles to treat the complicated granular powder materials. For ideal gas systems, the pressure is generated from the kinetic movements of gas molecules. To keep self-consistency, the pressure generated from granular powders should be analogously treated as resulted from the kinetic energy of particles. Eq. (12) is obtained under such an assumption with the consistent kinetic energy connection approach, rather than simply replacing the usual pressure with Janssen's equation. I have to say that Eq. (12) is not obtained with uncertain assumptions; rather it is derived with the same kinetic energy connection $\frac{3}{2} k_B T = \frac{1}{2} m v_{rms}^2$ that is the foundation of thermodynamics and has been used in granular powder systems for many years. Someone may argue that, according to Eq. (10) a good definition of granular temperature should vanish for an immobile packing of particles. However, per Eq. (9) the pressure is induced from the kinetic energy and we also know for sure that the pressure in granular powders is not zero. For keeping consistent with the granular temperature definitions using the kinetic energy approach, I have no choice but to come up Eq. (12), though the obtained granular temperature should be very close to zero.

The granular temperatures defined in Eq. (21), (24), and (30) are obtained under dynamic rather than stationary conditions. Eq. (21) is for granular powders under a continuous vibration shear, Eq. (24) is for granular powders under a continuous simple shear, and Eq. (30) is for the granular powders continuously rolling on a slope. The granular temperatures are defined on the continuous energy inputs into the systems, thus the time should be a critical parameter in the definition. After the granular temperatures are defined analogously with thermodynamics using the kinetic energy connection approach, particles may represent the very fundamental "atoms" of the thermal systems and thus thermodynamic principles can be applied to granular non-thermal systems. The goal of this article is to validate this approach and to see what predictions can be obtained. As indicated earlier, this approach is not originated by me and the success has been demonstrated in my recent four publications, ref. [30,31,62,70]. Since the granular temperatures are defined with the kinetic energy rather than the mean squared grain speed, the statistical framework is therefore not addressed in this article.

Although my article focuses on the jamming transition at dynamic conditions, the obtained equations are suitable for static jamming, too. Take Eq. (37) as an example, if there is no shearing and thus shear rate is zero, then the jamming volume fraction equals to the maximum packing fraction, which is very true at stationary conditions. The friction doesn't enter my description is because the jamming is defined as the immobility of particles due to the low kinetic energy or low granular temperature and the spatial crowdedness or the very small free volume in the systems. Microscopic frictional interactions are therefore not considered in my article.

As one may know, granular powders are athermal systems, utilization of thermodynamics and statistical mechanical theories extracted from conventional thermal systems to treat granular powders are found to be in good agreement with experimental results [12,28,62,70]. We thus examine the applicability of the four laws of thermodynamics on granular powder systems, and define granular temperatures for several granular powder systems in an analogous manner. The key point is to define the temperatures in granular powders through the kinetic energy connection with temperature, as shown in ideal gases. The main goal is to establish an approach that can facilitate easy applications of thermodynamic principles to granular powder systems. Such attempts have been made before for addressing both wet particle systems like colloidal suspensions and dry particle systems like granular powders. For examples, Hao [61] has successfully used the Eyring's rate theory [71] and the free volume concept for obtaining the viscosity equations of colloidal suspensions and polymeric systems with substantial modifications; A very similar theoretical approach is successfully employed to derive the two popular empirical tap density equations, the logarithmic and stretched exponential equations [62,70]. All these successes evidenced in literature imply that both thermal and athermal systems can be well described with common thermodynamic principles. What we need is a bridge that can build up a uniform connection between those two systems. This article represents the first attempt in this direction and further refinements are expected. Future attempts will be to utilize the Eyring's rate process theory and free volume concept to treat the granular systems for the purpose of deriving viscosity equations of granular systems under various conditions. Similar methods and approaches shown in our previous publications will be employed again to treat granular powder systems in a much natural manner, once the granular temperatures are properly defined.

## 6. Summary and conclusions

In summary, the thermodynamics originated from thermal systems is utilized to define the granular temperatures in granular systems in an analogous manner. The key point is to connect the kinetic energy to the temperature, and thus the temperature can be defined in a uniformed manner across the conventional thermal systems like colloidal suspensions to athermal systems like granular powders. This is a necessary step, as in granular systems thermal energy is too small to drive granular particle movements; new temperature definitions are needed for properly applying the thermodynamic principles established in thermal systems to granular systems. Several common granular systems are analyzed and the defined granular temperatures are summarized in Table 2. The obtained granular temperatures seem to be very high in comparison with the temperatures in thermal systems. However, please keep in mind that in conventional thermal systems, the molecule movements are very mild in much smaller distance scales; Lower temperatures seem to be adequate for thermal systems; On the other hand, the particle movements in granular powders are typically very intensive and wild, and higher granular temperatures seem to be adequate.

Once the granular temperatures are defined, the jamming temperature is analogously defined, too. The jamming particle volume fractions are thus obtained by assuming that the ratio of the granular temperatures to the jamming temperature equals to one. Therefore, the jamming points can be predicted and the obtained results agree qualitatively very well with experimental observations and empirical solutions in powder handlings. The particle volume fractions at jamming points obtained at several common cases are listed at Table 2.

The work in this article may lay a foundation for building up the "granular dynamics" on the basis of the granular temperatures defined analogously with that in thermodynamics. The four laws of thermodynamics are applicable to the granular powders with such definitions. Since the most important jamming phenomena in granular powders under a shear and a vibration are intensively examined, the results presented in this article may provide further insights on how to efficiently control the jamming process that has vast and important applications in industries like soft robotics and architecture [56].

**Acknowledgement:** The author appreciate Christine Heisler for reading through the manuscript and providing constructive comments.

Table 2 Proposed granular temperatures and the particle volume fractions at jamming points predicted in several granular systems.

| Granular systems | Granular temperatures | Typical values (K) | Conditions | Particle volume fractions at jamming points |
|---|---|---|---|---|
| Powders in stationary bins | $T_{gp} = \dfrac{\rho g D (2+K)}{16 n \mu K k_B} \left[1 - exp\left(-\dfrac{4\mu K h}{D}\right)\right]$ | Very close to zero | The particle number density n is a very large number | $\emptyset_j = \dfrac{\emptyset_m}{\left[1 + \dfrac{9D(2+K)}{96\mu K r \emptyset_A}\left[1 - exp\left(-\dfrac{4\mu K h}{D}\right)\right]\right]^3}$ |
| Powders in vibrations | $T_{gp} = \dfrac{8}{9} \dfrac{\pi r^3 \rho g L_0 n_v}{k_B}$ | $1.98 \times 10^{10}$ | r=1μm, ρ=1g/cm³, $L_o = 1\ cm$, $n_v = 1000$ | $\emptyset_j = \dfrac{\emptyset_m}{[1 + L_o \omega t/(2\pi r)]^3}$ |
| Powders under a shear | $T_{gp} = \dfrac{8}{9} \dfrac{\pi \sigma r^3 \dot{\gamma} \rho t}{k_B \rho_b}$ | $2.02 \times 10^{8}$ | r=1μm, ρ=1 g/cm³, $\rho_b = 0.3$ g/cm³, $\sigma = 1\ Pa$, $\dot{\gamma} = 1\ Hz$, $t = 300\ s$ | $\emptyset_j = \dfrac{\emptyset_m}{[1 + \sigma \dot{\gamma} t/(gr\rho_b)]^3}$ |
| Particles rolling on a slope | $T_{gp} = \dfrac{8\pi r^3 \rho}{9 k_B} g^2 (sin\theta - \mu cos\theta)^2 t^2$ | $4.76 \times 10^{11}$ | r=1μm, ρ=1 g/cm³, $\theta = 45°$, $t = 10\ s$, $\mu = 0.3$ | $\emptyset_j = \dfrac{\emptyset_m}{\left[1 + \dfrac{gt^2}{r}(sin\theta - \mu cos\theta)\right]^3}$ |